\documentclass[11pt,onecolumn,showpacs,superscriptaddress,preprintnumbers,floatfix,nofootinbib]{revtex4}

\usepackage{t1enc}
\usepackage{amssymb}
\usepackage{color}
\usepackage{graphics}
\usepackage[dvips]{graphicx}
\usepackage[cp1250]{inputenc}

\def\phiv{\overrightarrow{\phi}}
\def\calo{{\cal O}}
\def\eff{f_\phi}

\begin{document}

\title{On classicalization in nonlinear sigma models}
\bigskip

\author{R.~Percacci} 
\affiliation{
International School for Advanced Studies, 
via Bonomea 265, 34136 Trieste, Italy} 
\affiliation{INFN, Sezione di Trieste, Italy}
\author{L. Rachwa\l} 
\affiliation{
International School for Advanced Studies, via Bonomea 265, 34136 Trieste, Italy}
\affiliation{
The Abdus Salam International Centre for Theoretical Physics,\\
Strada Costiera 11, 34151 Trieste, Italy} 


\begin{abstract}
We consider the phenomenon of classicalization in nonlinear sigma models
with both positive and negative target space curvature and with any number of derivatives.
We find that the theories with only two derivatives exhibit a weak form
of classicalization, and that the quantitative results depend on the sign of the curvature.
Nonlinear sigma models with higher derivatives show a strong form of the phenomenon 
which is independent of the sign of curvature.
We argue that weak classicalization may actually be equivalent to asymptotic safety,
whereas strong classicalization seems to be a genuinely different phenomenon.
We also discuss possible ambiguities in the definition of the classical limit.
\end{abstract}
\maketitle

\section{Introduction}
The nonlinear sigma model and Einstein's theory of gravity show many similar features.
At the kinematical level, both theories have nonlinear configuration spaces,
which make their dynamics necessarily nonlinear too.
There is no ``zero field''
and the quantization is based on the use of the background field method.
In both cases the degrees of freedom can be viewed as Goldstone bosons
and their interactions involve derivatives
\footnote{See for example in \cite{percacci}.}.
Due to the nonpolynomial nature of the action,
it is natural to think of the fundamental fields as being dimensionless.
Aside from a vacuum term, the Lagrangian can be expanded as
\begin{equation}
\label{act1}
S=\sum_k\sum_n\bar g_{k,n}\calo_{k,n}\ ,
\end{equation}
where $\calo_{k,n}$ is an operator containing $k$ derivatives and $n$ powers of the fields. 
\footnote{Usually $k$ must be even.}
In natural units the coefficients $\bar g_{2,n}$ have dimension of mass squared 
and $\bar g_{4,n}$ are dimensionless.
In order to define a perturbative expansion with a canonically
normalized kinetic term, one usually redefines the fluctuation field
by a factor $\sqrt{g_{2,2}}=m$. Then one finds that the role of the perturbative coupling is
played by $1/m$. It has dimension of length, so these theories are power counting nonrenormalizable.
Perhaps more urgently, perturbative scattering amplitudes grow like powers
of momentum and exceed the unitarity bound for momenta comparable to $m$.
In fact, it is more correct to say that the perturbative expansion parameter is the dimensionless
ratio $p/m$, where $p$ is a typical momentum of the process under study,
so that the perturbative treatment is useful up to momenta of order $m$.
The standard view is then to regard these theories as effective field theories,
valid at energy and momentum scales below $m$.

In principle, however, it is possible that some of these theories may somehow heal
themselves of their perturbative problems.
One possibility is that the growth of the effective couplings such as
$p/m$ terminates due to quantum effects and that approaches a 
fixed point in the UV \cite{wilson}.
In particle physics and gravity this behavior is called ``asymptotic safety'' \cite{weinberg}.
There is by now significant evidence for the existence of asymptotically safe
RG trajectories in gravity, see for example \cite{reviews};
some work has also been done for the nonlinear sigma models \cite{codello,zanusso}
and in particular for the electroweak chiral model \cite{bfptvz}.
One expects that in such asymptotically safe theories
the scattering amplitudes also stop growing and respect the unitarity bounds,
although no complete calculation of this type has been performed so far.

More recently, a different idea has been proposed, namely that the
growth of the scattering amplitudes is controlled by the formation of classical intermediate states. 
In this picture, which has been called ``classicalization'',
a high energy quantum state with low occupation number
evolves into a classical state (called a ``classicalon'') when the radius
becomes comparable to a characteristic radius $r_*$ called ``classicalization radius''.
The important point is that $r_*$ does not decrease with energy as
one might naively think, but rather grows with it or at least tends to a constant.
We will call these cases ``strong'' and ``weak'' classicalization, respectively.
As a result, when the energy of the incoming states becomes greater than the
characteristic scale $m$, scattering is dominated by the formation of classicalons
and the cross section  tends to the geometrical value $r_*^2$.
The idea of classicalization emerged first in the case of gravity,
where the classicalons would correspond to black holes \cite{dvaligravity},
but subsequently it has been recognized as a possible behavior also in 
Goldstone bosons models \cite{dvali1,dvali2,dvali3,dvali4}.
Other aspects have been considered in \cite{bajc,akhoury,tetradis,	grojean}.

In spite of the evident differences between asymptotic safety and classicalization,
one wonders whether they might not be two ways of looking at the same phenomenon.
If, for example, the amplitude for Goldstone boson scattering unitarizes
at high energy without having to introduce new weakly coupled degrees of freedom,
it would be surprising if there existed two independent mechanisms by which Nature
could achieve this. If two explanations are available, they might just be
different descriptions of the same phenomenon.

Motivated in part by this question,
in this paper we discuss aspects of classicalization in the nonlinear sigma models.
We extend previous analyses by considering the effect of 
the curvature of the target space.
Much of the work that had been done previously had concentrated on a simple
model of a single scalar, and since a one dimensional space is flat,
interactions necessarily involve terms with more than two derivatives.
When the target space is curved, there are infinitely many
interaction terms already at the two derivative level.
We analyse the effect of these terms first by themselves,
and then in the presence of higher derivative interactions.
In order to be able to discriminate the effect of positive and negative curvature we
shall consider both spherical and hyperbolic target spaces.

If one wants to compare classicalization to asymptotic safety,
the first obvious difference is the fact that asymptotic safety is based on
renormalization group running, which is a quantum effect, 
whereas classicalization, as the name suggests seems to originate
from the formation of classical states.
We will argue in the last section that the distinction between quantum 
and classical phenomena is actually subject to various ambiguities in what
one really means by ``classical limit''.
Using a specific definition of a classical limit, one can 
view classicalization as originating from quantum effects.
Moreover we will show that weak classicalization
is a quantum effect and may be equivalent to asymptotic safety.
In order to disentangle classical from quantum effects we will work
throughout in units where $\hbar$ is not set equal to one.

We conclude this introduction by outlining the content of the following sections.
In section 2 we review the notion of classicalization in the case of a theory of a single
Goldstone boson with arbitrary derivative interactions.
In section 3 we discuss nonlinear sigma models with values in maximally symmetric spaces
with both positive and negative curvature, and with two derivatives only.
We find that a weak form of classicalization happens.
In section 4 we extend the analysis to include higher derivative terms.
In section 5 we return to the comparison between classicalization and
asymptotic safety and we draw our conclusions. 

\section{A single self-interacting Goldstone boson}  

In this section we begin by considering a model of a single Goldstone boson
with higher derivative interaction lagrangian of the form:
\begin{equation}
\label{lder}
 \mathcal{L}=\frac{1}{2} (\partial \phi)^2
 +\frac{L_*^{4(m-1)}}{2m} \left(\left(\partial \phi \right)^2 \right)^m\ .
\end{equation}
Here $m$ is an index counting the derivatives.
The field $\phi$ has the canonical dimension $M^{1/2}L^{-1/2}$ and the coupling $L_*$
has dimension $L^{3/4}M^{-1/4}$.
Later we will comment on the effect on classicalization of the presence of terms with 
lower or higher number of derivatives, but for the moment we assume that (\ref{lder}), 
with a fixed $m$, is the only interaction. 
The equation of motion coming from this lagrangian is
\begin{equation}
\label{eom}
\Box \phi 
+L_*^{4(m-1)}\partial^\mu \left[\partial_\mu \phi \left((\partial \phi)^2 \right)^{m-1}\right] =0\ .
\end{equation}
Assuming that free asymptotic states
solving the equation $\Box \phi_0=0$ exist,
the solution of the nonlinear equation (\ref{eom}) can be constructed perturbatively.
We consider solutions with spherical symmetry
\footnote{Then the divergence of a one-form $v_\mu$ is $\partial_0 v_0-\frac{1}{r^2} \partial_r (r^2 v_r)$
and the d'Alembertian is
$\Box =\partial_t^2 - \frac{1}{r^2} \partial_r (r^2 \partial_r)$}.
The initial ingoing unperturbed free wave has the form $\phi_0(t,r)=\sqrt{\hbar}\psi(\omega(t+r))/r$,
where $\psi(z)=A\sin(z)+B\cos(z)$ is dimensionless.
We will assume that the wavelength $\omega^{-1}$ is small compared to the radius $r$,
so that we can think of the solution as a harmonic function with a slowly-varying
$r$-dependent amplitude. At large distances the effect of the interaction is negligible 
(because of higher $1/r$ dependence of the interaction terms). 

The equation for the first order perturbation $\phi_1$ is
\begin{eqnarray}
\label{pert}
&&\left(1+L_*^{4(m-1)}((\partial\phi_0)^2)^{m-1}\right)\Box \phi_1
\nonumber\\
&&+2(m-1)L_*^{4(m-1)}((\partial\phi_0)^2)^{m-2}
\left(\partial^\mu\phi_0\partial^\nu\phi_0\partial_\mu\partial_\nu\phi_1+
2\partial_\nu\phi_0\partial^\nu\partial^\mu\phi_0\partial_\mu\phi_1\right)
\nonumber
\\
&&=-2(m-1)L_*^{4(m-1)}((\partial\phi_0)^2)^{m-2}
\partial^\mu\phi_0\partial^\nu\phi_0\partial_\mu\partial_\nu\phi_0 \ .
\end{eqnarray}
We have written on the left hand side of the equation
all the terms that contain derivatives of $\phi_1$
and on the right a source term containing only $\phi_0$.
This equation is still quite complicated.
However, we will see {\it a posteriori} that for the values of $r$ that we are interested in,
the terms on the l.h.s. that come from the interaction are small relative to $\Box\phi_1$.
For our purposes it will therefore be sufficient to retain in the l.h.s. only the term $\Box\phi_1$.
\footnote{Additionally one could observe that as long as $\phi_1$ is a small perturbation
relative to $\phi_0$, the terms on the l.h.s. coming from the interactions must be small
relative to the source term on the r.h.s. .}

We make an ansatz 
\begin{equation}
\label{ansatz}
\phi_1(t,r)=\sqrt{\hbar}f(r)\eta(\omega(t+r))\ .
\end{equation}
In the approximation $\omega r\gg1$ we have
\begin{equation}
\Box\phi_1
\simeq -\frac{2\omega\sqrt{\hbar}}{r} \eta' \left(f r \right)'\ ,
\end{equation}
where a prime denotes derivative of a function with respect to its argument.
Then the equation for $\phi_1$ in the leading order of our approximation is
\begin{equation}
\label{leading1}
-\frac{2\omega\sqrt{\hbar}}{r} \eta' \left(f r \right)' = 
- \frac{2^{m-1} (m-1) L_*^{4(m-1)}\omega^m(\sqrt{\hbar})^{2m-1}}{r^{3m-1}}
\psi^{m-1}\psi'^{m-2}\left[\psi \psi'' +4\psi'^2\right]\ .
\end{equation}
The solution of this equation can be expressed as
\begin{equation}
\phi_1 = -\frac{2^{m-1} L_*^{4(m-1)}E^{m-1}\sqrt{\hbar}}{6r^{3m-2}} \eta(\omega(t+r)),
\end{equation}
where $E=\hbar\omega$ and 
$\eta(z)=\int^z \psi^{m-1}\psi'^{m-2}\left[\psi\psi'' +4\psi'^2\right] dz'$.
Note that for any $m$ the integrand is an odd and periodic function with period $2\pi$
and such that the integral over one period is zero.
Therefore the function $\eta$ is again dimensionless and periodic with period $2\pi$,
which means that the scattered wave has the same frequency as the incoming one.

Since $\eta\sim\psi\sim 1$, 
the ratio of the amplitudes of the first perturbation to the initial wave can be expressed as
\begin{equation}
|f(r) r|\simeq \frac{L_*^{4(m-1)}2^{m-1} E^{m-1}}{r^{3(m-1)}} = \left(\frac{r_*}{r} \right)^{3(m-1)}\ ,
\end{equation}
where in the last step we defined the ``classicalization radius'' $r_*=\sqrt[3]{2L_*^4 E}$. 
Notice that it does not depend on $m$.
We can now see why the interaction terms on the l.h.s. of (\ref{pert}) are negligible.
For example the second term in the first bracket is of order $(E L_*^4/r^3)^{m-1}$,
and when $r\gg r_*$ we have $(EL_*^4/r^3)\ll1$.
Similar considerations apply to the other terms.

We thus find that the scattering process becomes important at distances of order $r_*$.
Normally one would expect a scattering  process involving particles with energy $E$
to probe distances of order $\omega^{-1}=\hbar/E$, but since $E>\hbar^{3/4}/L_*$ implies $r_*>\omega^{-1}$,
we find that for energies exceeding the characteristic energy scale of the model $\hbar^{3/4}/L_*$,
the distances probed actually increase with energy.
We will call this behavior strong ``classicalization''.

The non-spherically symmetric case has been discussed in \cite{akhoury}.
For mild deformations, it was found that the classicalization radius
becomes smaller (larger) in regions where the curvature of the incoming wave
is smaller (larger). 
Since the preceding arguments were order-of-magnitude estimates anyway,
this does not change the conclusions.
In the limiting case when the incoming wavefronts are flat,
the classicalization radius goes to zero and hence no classicalization occurs.

Let us now allow for the simultaneous presence of the interaction terms with different values of $m$.
Motivated by effective field theory, we assume that all interactions
are of the form
\begin{equation}
 \mathcal{L}_{\textrm{int}}=\sum_m c_m L_*^{4(m-1)} \left(\left(\partial \phi \right)^2 \right)^m
\label{lint}
\end{equation}
To each interaction there corresponds a classicalization radius given by
$r_*^3 =  2\sqrt[m-1]{2m c_m} EL_*^4$. 
Which one of these scales plays the dominant role depends on the 
dimensionless coefficients $c_m$.
If $c_m\sim1/m$, as we assumed earlier, they are all of the same order and therefore in principle
all terms in the Lagrangian are equally important.
On the other hand if $4 c_2 > \sqrt{6 c_3 } > \sqrt[3]{8 c_4}>\ldots$,
then the corresponding $r_*$ decreases with $m$, and the four-derivative term is the 
most important one.
For large $m$ one could assume that the coefficients $c_m$ do not grow faster 
than exponentials of $m$ ($c_m < a^m/2m $ for some $a>1$).
This condition is quite reasonable for effective field theories.
Under these conditions the system will classicalize when its size reaches
the largest of all these possible classicalization radii
and the higher derivative interactions will not play any significant role.

\section{Nonlinear sigma model with 2 derivatives}
When there is more than one Goldstone boson, the theory admits interaction terms
with just two derivatives. 
A standard way of describing the dynamics is to package the kinetic and
the two-derivative interaction terms in the geometrical form
\begin{equation}
\mathcal{L}=\frac{1}{2} h_{ab}\partial_\mu\phi^a\partial^\mu\phi^b ,
\label{sigmaL}
\end{equation}
where $h_{ab}$ is a metric in the target space depending on $\phi$ .
The coefficients of the Taylor expansion of the metric around a constant $\phi$
can be viewed as an infinite set of coupling constants.
We will consider real, maximally symmetric target spaces,
for which all couplings are related and the only free parameter is the overall scale of the metric.
There exist coordinates such that 
\begin{equation}
h_{ab} = \delta_{ab}\pm \frac{\phi^a\phi^b}{\eff^2\mp\phiv^2}\ ,
\end{equation}
where the $+$ and $-$ signs correspond to positive and negative curvature
(sphere and hyperboloid). 
In the above formula $\eff$, which has the same dimensions as the field,
has the meaning of radius of the sphere or hyperboloid in field space 
and $\phiv^2=\phiv \cdot \phiv=\delta_{ab} \phi^a \phi^b$ is the usual flat Euclidean product. 
Lorentz indices will be suppressed when this doesn't lead to confusion.
Using the explicit form of the metric, the lagrangian (\ref{sigmaL}) can be put in the form
\begin{equation}
\label{nlsm}
\mathcal{L}=\frac{1}{2}\left[(\partial \phiv)^2 
\pm \frac{(\phiv \cdot \partial\phiv)^2}{\eff^2\mp\phiv^2}\right]\ .
\end{equation}
The equations of motion are
\begin{equation}
\Box \phi^a \pm \frac{\phi^a\,\partial(\phiv \cdot \partial \phiv) }{\eff^2\mp\phiv^2} 
\pm \frac{\phi^a\,(\phiv \cdot \partial \phiv)^2 }{(\eff^2\mp\phiv^2)^2} = 0\ .
\end{equation}

As in the preceding section, we are going to look for perturbative solution in the form $\phiv=\phiv_0+\phiv_1+\ldots$, where $\phiv_0$ is a solution of the free wave equation: $\Box \phiv_0=0$.
We will study to which extent in spacetime evolution we can treat $\phiv_1$ as a small perturbation solving approximately the interacting field equations.
In order to follow more closely the analysis of the preceding section,
it is tempting to try and reduce the problem to a single-field problem by assuming
that only one component of the field is nonzero.
The equations of motion seem to retain much of their nonlinearity even in this case.
This, however, is an illusion that can be easily undone by a field redefinition.
For example, with a single-field ansatz the Lagrangian (\ref{nlsm}) becomes
\begin{equation}
\frac{1}{2} \left(\partial\phi\right)^2\frac{\eff^2}{\eff^2\mp\phi^2}
\end{equation}
and this can be recast as a free field Lagrangian by the redefinition $\phi=\eff\sin\varphi$
(for the upper sign) or $\phi=\eff\sinh\varphi$ (for the lower sign).
This means that if we make a single-field ansatz we will not be able to
detect effects due to curvature, which is one of our purposes.
Without much loss of generality we will work with a general spherically symmetric incoming wave
$\phi_0^a(r,t)=\sqrt{\hbar}\psi_a(\omega(t+r))/r$, where all components have the same frequency,
and we assume $\omega r\gg1$, as before.
The first order perturbation will be written in the form
$\phi_1^a(r,t)=\sqrt{\hbar}\eta_a(\omega(t+r))f(r)$. 
Later we will see that it is consistent to assume that all components of $\phi_1^a$
have the same radial dependence.

Linearizing the field equation around $\phiv_0$ we find
\begin{eqnarray}
&&h_{ab}\Box\phi_1^b 
\pm \frac{2\phi_0^a}{\eff^2-\phiv_0^2} h_{bc}\partial\phi_0^b\partial\phi_1^c
\pm \frac{\phi_1^a}{\eff^2-\phiv_0^2} h_{bc}\partial\phi_0^b\partial\phi_0^c
\nonumber
\\
&&+ \frac{2\phi_0^a}{(\eff^2-\phiv_0^2)^2} 
\left[(\partial\phiv_0)^2\phi_0^b+(\phiv_0\partial^\mu\phiv_0)\partial_\mu\phi_0^b
\pm2\frac{(\phiv_0\partial\phiv_0)^2}{\eff^2\mp\phiv_0^2}\phi_0^b
\right]
\phi_1^b
\nonumber\\
&&
=
\mp\frac{\phi_0^a}{\eff^2-\phiv_0^2} h_{bc}\partial\phi_0^b\partial\phi_0^c\ .
\label{scalareom}
\end{eqnarray}
Here the metric $h_{ab}$ has to be regarded as a function of $\phiv_0$.
As in the preceding section, the interaction terms on the l.h.s. can be neglected.
To leading order in $1/r\omega$ we find 
\begin{equation}
\label{elena}
-\frac{2\omega\sqrt{\hbar}}{r}(f r)'\eta_a' = 
\mp \frac{2\omega\hbar^{3/2}}{\eff^2r^4}
\frac{\psi_a\vec\psi\vec\psi'}{\left(1\mp\frac{\hbar\vec\psi^2}{\eff^2 r^2}\right)^2}\ .
\end{equation}
We note right away that in contrast to equation (\ref{leading1}) the $\omega$-dependence
will cancel out. Instead, the behavior of the solution is governed by the new dimensionless parameter 
$\eff r/\sqrt{\hbar}$.
As long as $\eff r/\sqrt{\hbar}\gg1$, the denominator in the r.h.s. can be approximated by one
and the equation can be solved by separation of variables. Now we can notice that
after separation the radial equation for $f$ is the same for all components of $\phi_1^a$,
therefore the choice $f^a(r) = f(r)$ is justified.
The solution can be written in the form
\begin{equation}
\label{george}
\phi_1^a = \mp \sqrt{\hbar}\frac{\hbar}{2 \eff^2 r^3} \eta_a(\omega(t+r))\ ,
\end{equation}
where
$\eta_a(z)=\int^z \psi^a \overrightarrow{\psi} \overrightarrow{\psi}' dz'$.
This is again an oscillating function with $r$-dependent amplitude,
but in contrast to the case of the preceding section, the amplitude of the oscillations
of the scattered wave is independent of $\omega$.  
The ratio between the amplitude of the first perturbation and the incoming wave is
\begin{equation}
\label{compare}
\left| f(r) r\right| =  \frac{\hbar}{2 \eff^2 r^2}=\left( \frac{r_*}{r}\right)^2\ .
\end{equation}
From the above expression we see that we can define a ``classicalization radius''
\begin{equation}
r_*=\frac{\sqrt{\hbar}}{\sqrt{2}\eff}
\end{equation}
independent on the frequency or energy of the incoming wave packet.
Again, incoming waves with arbitrarily high frequency are unable to probe
distances shorter than $r_*$, but in contrast to the preceding case $r_*$
does not increase with frequency. 
We thus have a weaker form of classicalization (compare \cite{dvali4}).

Let us now consider the effect of curvature, which (aside from the immaterial
overall sign) is contained in the denominator of the r.h.s. of (\ref{elena}).
We observe that since $0\leq \vec\psi^2 \leq C$, for some constant $C$ of order one,
the effect of the denominator is to enhance the amplitude of the scattered wave
for positive curvature (upper sign) and to decrease it for negative curvature (lower sign).
In fact, with the positive curvature the amplitude reaches a pole for some $r\approx \sqrt{\hbar}/\eff$,
strengthening the case for classicalization of the preceding analysis.
In the case of negative curvature, rhe r.h.s. of (\ref{elena}) increases for
decreasing radius and tends to a constant for $r\to 0$.
The argument for classicalization is considerably weaker in this case.

This can also be seen in another way. The approximation leading to (\ref{george})
corresponds to considering the theory with interaction Lagrangian
\begin{equation}
 \mathcal{L}_{\textrm{int}} = \pm \frac{(\phiv \cdot \partial \phiv)^2}{2\eff^2}\ .
\end{equation}
Let us consider what happens if we take as an interaction the next term in the
expansion of the denominator
\begin{equation}
 \mathcal{L}_{\textrm{int}} = - \frac{\phiv^2(\phiv \cdot \partial \phiv)^2}{2\eff^4}\ .
\end{equation}
From here one finds instead of (\ref{elena}) the equation
\begin{equation}
-\frac{2\omega\sqrt{\hbar}}{r}(f r)'\eta_a' = 
- \frac{2\omega\hbar^{5/2}}{\eff^4r^6}
\psi_a\vec\psi^2\vec\psi\vec\psi'\ ,
\end{equation}
whose solution has a radial dependence
\begin{equation}
\left| f(r) r\right| =  \frac{\hbar^2}{2\eff^4 r^4}\ .
\end{equation}
This corresponds again to a classicalization radius of order $\sqrt{\hbar}/\eff$.
It is easy to see that this is true for all the terms in the expansion,
but when one takes them all into account simultaneously,
they appear all with negative sign when the curvature is positive,
but with alternating signs when the curvature is negative.

In the case of an incoming plane wave,
the ratio of the first perturbation to the initial amplitude 
is independent both of $\omega$ and $r$. 
This gives no clue about classicalization. 
These considerations in the planar wave case seem to confirm the previous statement, that if 
classicalization holds for NSM with 2 derivatives, it is in the weak form.

\section{Nonlinear sigma model with 2 and 4 derivatives}

In a maximally symmetric nonlinear sigma model with a two-derivative Lagrangian (\ref{nlsm}),
a general four derivative interaction has the form
$$
\mathcal{L}^{(4)}_{\textrm{int}} =  
g_4
(\ell_1 h_{ab} h_{cd}+ \ell_2 h_{ac} h_{bd})
\partial_\mu\phi^a \partial^\mu\phi^b \partial_\nu\phi^c \partial^\nu\phi^d\ ,
$$ 
where $\ell_1$ and $\ell_2$ are dimensionless constants of order one.
Expanding the metrics in Taylor series would yield infinitely many
monomial operators with coefficients $g_{4,n}$.
For the sake of comparison to section II we could write $g_4=L_*^4$.
In effective field theory one expects the coefficients of operators
with different number of derivatives to be all proportional to powers
of the same mass scale $\eff$ in natural units. Then we would write alternatively
$g_4=\hbar/\eff^4$. 
We will follow this notation here, but one can revert to $L_*$ at any moment.

When this interaction is added to the two-derivative Lagrangian (\ref{nlsm}),
applying the same ansatz for the fields as in the preceding section,
neglecting $\phiv_1$ on the l.h.s. and 
expanding in inverse powers of $\omega r$ we get to the leading order
the following equation for the perturbation:
\begin{eqnarray}
\label{johnny}
&&\Box \phi_1^a =
\mp \frac{2\omega\hbar^{3/2}}{\eff^2r^4}
\frac{\psi_a\vec\psi\vec\psi'}{\left(1\mp\frac{\hbar\vec\psi^2}{\eff^2 r^2}\right)^2}
\\
&& 
- \frac{2\omega^2\hbar^{5/2}}{\eff^4 r^5}
\left[(\ell_1+3\ell_2)\psi_a\vec\psi^{\prime2}
+(3\ell_1+5\ell_2)\psi'_a\vec\psi'\vec\psi
+(\ell_1+\ell_2)\psi_a\vec\psi''\vec\psi 
+\ell_2\psi''_a\vec\psi^2\right]
\nonumber
\end{eqnarray}
Note that in the four-derivative terms the $\phi$-dependent part of the metric
gives subleading contributions, so $h_{ab}$ has been already replaced by $\delta_{ab}$ in (\ref{johnny}).

The equation can only be solved by separation of variables if one of the two terms in (\ref{johnny})
can be neglected. 
However, we can get a reasonably good estimate of the terms involved by simply
setting equal to one factor $\eta$ in the l.h.s. and all the terms involving
$\psi$ in the r.h.s..
The resulting equation for $f(r)r$ can then be easily integrated to yield
\begin{equation}
|f(r)r| =
\mp\frac{\hbar}{2\eff^2r^2}
-\frac{E\hbar}{3\eff^4r^3}
-\frac{\hbar^2}{4\eff^4 r^4}+\ldots
\,
=\mp\left(\frac{r_{2*}}{r}\right)^2
-\left(\frac{r_{4*}}{r}\right)^3
-\left(\frac{r_{2*}}{r}\right)^4+\ldots
\end{equation}
where the first and third term come from the expansion of the two-derivative term
and the second comes from the four-derivative term.
We have defined the classicalization radii 
\begin{equation}
r_{2*}=\sqrt{\frac{\hbar}{2\eff^2}}\ ;\qquad
r_{4*}=\sqrt[3]{\frac{E\hbar}{3\eff^4}}\ .
\end{equation}
All the terms in the expansion of the two-derivative term correspond to the
same classicalization radius $r_{2*}$.
These terms are dominant for $E<\sqrt{\hbar}\eff$.
For higher energy the four-derivative terms dominate and the system
behaves like $N$ copies of the single Goldstone boson model of section II,
in the special case with $2m=4$ derivatives. 
Note that if we use the notation  $L_*^4=\hbar/\eff^4$,
$r_{4*}=\sqrt[3]{EL_*^4/3}$, which is the same formula that we found in section II.
Strong classicalization occurs for $\omega > r_{4*}^{-1}$ regardless of the sign of the curvature.

In the case of a plane incoming wave we also have to distinguish two regimes.
When the two-derivative terms in (\ref{johnny}) dominate, 
no clues of classicalization can be found, as in section III.
When the four-derivative term dominates classicalization does not occur,
in agreement with the discussion in section II and with \cite{akhoury}.

\section{Classicalization vs. asymptotic safety}

In the preceding sections we have analyzed a hypothetical scattering process
in nonlinear sigma models with any number of derivatives and with positive,
negative or zero target space curvature.
We have found that quite generally, an incoming spherical wave satisfying the
free wave equation will generate a strong scattered wave when it reaches a size $r_*$
that depends in general on the couplings of the theory and on the initial energy.
Contrary to naive expectation, this radius $r_*$ either increases with energy
or is independent of it.
As discussed in \cite{dvali2}, this is in sharp contrast to other field theories
such as a scalar with a potential interaction, where the scattered wave only becomes
important at a radius of order $\hbar/E$.
Following \cite{dvali1,dvali2,dvali3,dvali4}, we call this phenomenon
``classicalization'', and for our purposes we distinguish a 
``weak classicalization'', when $r_*$ is independent of $E$,
from ``strong classicalization'' when $r_*$ grows with $E$.
In both cases scattering processes cannot actually probe distances shorter than $r_*$.
The scattering process is softened and there is a chance that, though perturbatively
nonrenormalizable, the theory may actually be well behaved at high energy.

As already mentioned in the introduction, this sounds sufficiently similar to the
program of asymptotic safety that one may legitimately ask whether there is
a relation between the two phenomena.
To further motivate this expectation, let us recall that in order to
avoid the complications due to redundant (or ``inessential'') couplings,
in the discussion of asymptotic safety
it would be desirable to define the couplings directly in terms of physical
observables \cite{weinberg}.
Due to the difficulty of nonperturbatively computing observables in these theories,
so far efforts have concentrated on the running of couplings
defined as coefficients of operators in an effective Lagrangian.
However, if there was a way of showing, for example, that certain amplitudes have the right
behavior as functions of energy, then one could show that the couplings defined 
in terms of the corresponding exclusive cross sections would reach a fixed point.
In this way classicalization could turn out to be a valuable alternative tool
for studying asymptotic safety.

Since asymptotic safety, if realized in nature, is clearly a quantum phenomenon,
the first priority is to understand whether there is a way of viewing also classicalization
as a quantum phenomenon, in spite of its name.
We believe that the distinction between classical and quantum phenomena
is not as clear cut as it seems.
The real world is quantum in nature and classical behavior can only emerge
in certain limits, but there are ambiguities in the way these limits are taken.
We refer to \cite{brodsky} for a recent discussion of this issue in the context of QED.
In order to introduce the issue in the context of the nonlinear sigma model,
let us go back to the parametrization where the fields $\varphi^a$ are dimensionless
(which is natural in view of the fact that they appear as arguments in nonpolynomial functions).
The action can be expanded schematically as in (\ref{act1})
where ${\cal O}_{k,n}\sim\int\partial^k\varphi^n$ contains $k$ derivatives
and $n$ powers of the field.
The dimensions of the couplings $\bar g_{k,n}$ are $ML^{k-3}$, independent of $n$.
For the sake of perturbation theory one has to separate the kinetic term from interactions.
Defining a canonically normalized field $\phi^a=\varphi^a\sqrt{g_{2,2}}$, 
of dimension $\sqrt{M/L}$, the action becomes
\begin{equation}
\label{act2}
S=\int\left[(\partial\phi)^2+\sum_k\sum_{n>2} g_{k,n}\partial^k\phi^n\right]
\end{equation}
where $g_{k,n}=\bar g_{k,n}(\sqrt{g_{2,2}})^n$ have dimension $M^{1-n/2}L^{k-3+n/2}$.
There is a theorem to the effect that higher derivative corrections to the propagator
can be eliminated by field redefinitions, order by order in perturbation theory \cite{anselmi},
so we may assume without loss of generality that $g_{k,2}=0$ for $k>2$.
Assuming that a $Z_2$ symmetry forbids the appearance of odd powers of the field,
the lowest interaction would be of the form $g_{2,4}\phi^2(\partial\phi)^2$.
Let us define $g_{2,4}=\eff^{-2}$, where $\eff$ has the same dimensions as the field
(it can be viewed as a kind of VEV).
Global symmetry then implies that $g_{2,n}\sim\eff^{2-n}$ (see for example (\ref{nlsm})).
In effective field theory it seems reasonable to assume that all dimensionful couplings
are proportional to powers of $\eff$. 
(This is particularly clear in natural units, where $\eff$ can be viewed as a natural
mass scale, and all couplings are proportional to powers of this mass.)
Then we would write $g_{k,n}=c_{k,n} \eff^{4-k-n}\hbar^{k/2-1}$,
where $c_{k,n}$ are dimensionless.

One can define different notions of classical limit,
depending on which couplings are being kept fixed.
If one takes $\hbar\to0$ keeping $g_{k,n}$ fixed,
one obtains a classical field theory with all the higher derivative terms;
if one takes $\hbar\to0$ keeping $\eff$ and the $c_{k,n}$ fixed one gets 
a classical field theory with the two-derivative terms only.
How one defines the classical limit obviously affects the interpretation of classicalization.
In the former limit the classicalization radius, when $k>2$, is
$(g_{k,n})^{\frac{2}{n+2k-6}}E^{\frac{n-2}{n+2k-6}}$ independent of $\hbar$
and is therefore a truly classical notion \cite{dvali1}. 
In the latter limit, reexpressing $g_{k,n}$ in terms of $\eff$ and $\hbar$,
the classicalization radius goes to zero and should therefore be regarded as a quantum effect.
The classicalization radius found in section III, for the case $k=2$, is 
truly of quantum nature regardless which limit is taken.

Another potential source of ambiguity in the definition of the classical limit
is the question whether $E$ or $\omega$ is to be held fixed \cite{brodsky}.
In the latter case again the classicalization radius $\sqrt[3]{L_*^4 \hbar \omega}$
vanishes in the classical limit.
Since in this paper we are mainly interested in scattering experiments
where the momenta of the external particles are known and fixed,
it seems more appropriate to stick to the case when $E$ is kept fixed in the classical limit.
Furthermore, writing the couplings in terms of powers of a single coupling $\eff$ is motivated
by perturbative arguments. Since both asymptotic safety and classicalization
are nonperturbative notions, it is perhaps more appropriate to stick to the generic
parameterization (\ref{act2}) and to consider all couplings $g_{k,n}$ as truly independent.
This is the notion of classical limit which is implicitly assumed in 
\cite{dvali1,dvali2,dvali3,dvali4}.

We now restrict ourselves to this particular notion of classical limit,
and we try to extract some conclusions from the results of the preceding sections.
From the given expressions for $r_*$ we see that the weak classicalization
that was found in the two-derivative models of section III is a quantum phenomenon,
whereas the strong classicalization of the higher derivative models
of sections II and IV are genuinely classical effects.
There is therefore a chance that weak classicalization has something to do
with asymptotic safety, whereas strong classicalization seems to be a genuinely
different effect \footnote{Perhaps the names are not appropriate here.}.
There are then some other suggestive facts.
It was found in \cite{codello} that in the two-derivative truncation of the
nonlinear sigma model a non-trivial fixed point exists for positive curvature
but not for negative curvature.
This seems to agree with the result in section III, according to which
the argument for (weak) classicalization is much more robust in the
positive curvature case than in the negative curvature case.
On the other hand, no non-trivial fixed point seems to exist in the
$S^1$-valued nonlinear sigma model, which corresponds to the single
Goldstone boson model of section II \cite{zanusso}.
And furthermore, we have found in section IV that strong classicalization
is completely insensitive to the sign of the curvature.
Finally, returning to natural units,
the amplitude for scattering of two particles 
into two particles in the two-derivative model behaves like $p^2/\eff^2$,
where $p$ is the momentum transfer.
Since the latter is asymptotically of order $r_*^{-1}\sim \eff$, the amplitude
tends to a constant, as one would expect in an asymptotically safe theory.

In the case of gravity, it has been argued that classicalization is intimately
related to the notion of a minimal length \cite{dvaligravity}.
This seems to be in contrast to the notion of a field theoretic UV completion,
where one talks of ``arbitrarily high energy scales''.
In fact it had already been noted that in a certain sense a notion of minimal length
is present in an asymptotically safe theory of gravity \cite{perini3}.
We refer to \cite{vacca} for further discussion of this point.

All these facts reinforce the hypothesis that weak classicalization may be a 
direct manifestation of asymptotic safety in the scattering amplitudes
whereas strong classicalization, if true, would be a different kind of effect.
We also observe that if we assume equivalence between weak classicalization
and asymptotic safety, the absence of classicalization in the case of
plane waves suggests that momentum transfer is more important than total
energy in these matters.
In order to substantiate the preceding conclusions one would need to
directly calculate some amplitudes in an asymptotically safe theory.

\bigskip

\leftline{\bf Acknowledgements}
We would like to thank A. Kehagias for discussions.
This work was supported in part by the MIUR-PRIN contract 2009-KHZKRX. 
L.R. is supported by the European Programme {\it Unification
in the LHC Era} (UNILHC), under the contract PITN-GA-2009-237920.

\end{document}